\documentclass[twocolumn]{aastex701}

\usepackage{amsmath}
\usepackage{amssymb}
\usepackage{subcaption}
\usepackage{graphicx}
\usepackage{verbatim}

\newcommand{\target}{SDSS\,J1430$+$2303}

\newcommand{\kms}{\,\mathrm{km\,s}^{-1}}
\newcommand{\hst}{\textit{HST}}
\newcommand{\swift}{\textit{Swift}}
\newcommand{\xmm}{\textit{XMM-Newton}}
\newcommand{\nicer}{\textit{NICER}}
\newcommand{\civ}{\ion{C}{4}\,$\lambda$1549}
\newcommand{\mgii}{\ion{Mg}{2}\,$\lambda$2798}
\newcommand{\oiii}{[\ion{O}{3}]\,$\lambda$5007}
\newcommand{\hbeta}{H$\beta$}
\newcommand{\halpha}{H$\alpha$}

\shorttitle{HST Constraints on the Binary SMBH Candidate J1430+2303}
\shortauthors{Gross et al.}

\begin{document}

\title{\hst\ Ultraviolet Spectroscopic and Multi-band Imaging Constraints on the Candidate Coalescing Supermassive Black Hole Binary \target}

\correspondingauthor{Arran C. Gross}
\email{acgross@illinois.edu}

\author[0000-0001-7681-9213]{Arran C. Gross}
\affiliation{Department of Astronomy, University of Illinois at Urbana-Champaign,
1002 W. Green Street, Urbana, IL 61801, USA}
\email{acgross@illinois.edu}

\author[0000-0003-0049-5210]{Xin Liu}
\affiliation{Department of Astronomy, University of Illinois at Urbana-Champaign,
1002 W. Green Street, Urbana, IL 61801, USA}
\affiliation{National Center for Supercomputing Applications, University of Illinois at Urbana-Champaign,
1205 W. Clark Street, Urbana, IL 61801, USA}
\email{xinliul@illinois.edu}

\author[0000-0001-5769-0821]{Liam Nolan}
\affiliation{Department of Astronomy, University of Illinois at Urbana-Champaign,
1002 W. Green Street, Urbana, IL 61801, USA}
\email{liamjn@illinois.edu}

\begin{abstract}
\target\ is a low-redshift ($z=0.081$) Seyfert galaxy whose chirping optical and X-ray light curve over 2018--2022 was interpreted as a candidate uneven-mass-ratio highly eccentric supermassive black hole (SMBH) binary on the verge of coalescence \citep{Jiang2022arxiv}. We present \hst/STIS near-ultraviolet (UV) spectroscopy and \hst/WFC3 three-band imaging obtained to test that interpretation.  We address two largely independent predictions of the binary scenario. First, a spectral deficit (``notch'' or ``cutoff'') in the near-UV continuum expected from a circumbinary accretion disk: we find no coherent deficit, although systematics in modeling the UV Fe {\sc ii} emission preclude a quantitative upper limit. We caution that this null result is not, by itself, decisive against a binary, because general-relativistic magnetohydrodynamic (GRMHD) simulations including minidisks predict spectra that differ only modestly from single-AGN templates \citep{dAscoli2018}. Second, a kinematic test based on the broad emission lines: in a relativistic disk-emitter, the low-ionization \mgii\ line should track the multi-peaked Balmer profile, whereas in a binary in which one SMBH retains a bound broad-line region , \mgii\ and the broad Balmer lines should be similarly profiled and share a coherent centroid offset from \oiii. Both broad lines are single-peaked with matching widths and lack the double-peaked structure expected for a disk-emitter; the broad \mgii\ core is blueshifted relative to that of \hbeta\ by $\sim640 \kms$, a $\sim3\sigma$ offset once cross-grating systematics are included. Broad \civ\ is single-peaked and blueshifted, consistent with a normal accretion-disk wind and uninformative as a centroid anchor. WFC3 imaging shows a smooth elliptical/S0-like host with no significant tidal features at a depth of $\mu\sim$25.1 mag/arcsec$^{2}$. We interpret our \hst\ constraints in the context of post-2022 multi-wavelength follow-up \citep{Dotti2023,Marin2023,Masterson2023,Hoshi2024,An2026, He26}, which has found no continuation of the chirping periodicity and disfavors the imminent-merger interpretation.  

\end{abstract}


%-----------------------------------------------------------------
\section{Introduction}\label{sec:intro}

Supermassive black hole (SMBH) binaries are a generic prediction of hierarchical structure formation \citep{Begelman1980}.  After a major galaxy merger dynamical friction and stellar/gas-driven hardening deliver the central SMBHs to sub-parsec separations, beyond which gravitational radiation drives the inspiral to coalescence. These mergers are expected to dominate the low-frequency gravitational-wave sky accessible to pulsar-timing arrays (PTAs) and to future pace-based laser interferometers \citep{Sesana2017,Schutz2018}. Direct identification of a SMBH binary in the late inspiral would be transformative for multi-messenger astronomy and a critical complement to the stellar-mass mergers found by LIGO/Virgo/KAGRA \citep{Abbott2016}.

Despite theoretical expectations, no SMBH binary has been unambiguously confirmed at the gravitationally bound stage. More than 150 quasars have been proposed as candidates from apparent periodicities in optical light curves \citep{Valtonen2008,Graham2015,Charisi2016,Liu2019,Chen2020}. These claims remain controversial because quasar variability is intrinsically red noise dominated, which can mimic short-baseline periodicity \citep{Vaughan2016,Goyal2018,Zhu2020}. Spectroscopic searches for sub-parsec binaries through long-baseline radial-velocity monitoring of broad emission lines have likewise yielded only candidates and have constrained the bound-binary fraction at $\sim 0.1\,$pc separations to $\lesssim 1\%$ of optically active SMBHs \citep{Eracleous2012,Liu2014,Runnoe2017,Wang2017,Guo2019}. The most secure example of a periodic nuclear transient with a measured period derivative remains ASASSN-14ko, most plausibly a repeating partial tidal disruption event \citep[TDE; ][]{Payne2021,Payne2023}.

\subsection{The candidate SDSS J1430+2303}

\citet{Jiang2022arxiv} (hereafter J22) reported a remarkable optical and X-ray flare in the nearby ($z=0.0811$) Seyfert galaxy \target. The combined Zwicky Transient Facility (ZTF), \swift/XRT, and \swift/UVOT light curves show six oscillations over 2018--2022, with the period dropping from $\sim 1$\,year to $\sim 1$\,month and the amplitude decreasing in parallel. J22 model the chirping flare as an unequal-mass-ratio, highly eccentric SMBH binary in which the secondary punches through the primary's accretion disk near pericenter, and predict coalescence within roughly one to three years of the discovery epoch. We note that J22 itself remains an unrefereed preprint as of submission of this paper; the refereed multi-wavelength record on the original flare is provided by the \xmm/\nicer\ X-ray analysis of \citet{Dou2022}, the European VLBI Network observations of
\citet{An2022}, and the \swift/\xmm/\nicer\ follow-up of \citet{Masterson2023}.

The implied period derivative $\dot P$ of \target\ is striking. Adopting the peak times tabulated by J22, the optical flare-to-flare intervals shorten from 356 to 119\,days, and the optical-to-X-ray transition implies a final interval of $\sim 23$\,days. The mean $|\dot{P}|\sim 0.3$ on the optical baseline, and $\dot P$ approaching unity on the optical-to-X-ray transition, is two to three orders of magnitude larger than the $|\dot P|=(2.6\pm 0.6)\times 10^{-3}$ measured for ASASSN-14ko \citep{Payne2021}, where partial tidal disruption is the favored mechanism \citep{Cufari2022}. This very large $|\dot P|$ is inconsistent with gravitational-wave-driven inspiral for any plausible binary mass unless the system is within months of merger, leaving either a literal late-inspiral binary or a one-shot transient phenomenon as the two physically distinct possibilities.

\subsection{Post-2022 follow-up has weakened the binary case}\label{sec:status}

The predicted merger has not been observed, and a substantial body of follow-up observations has accumulated. The chirping periodicity did not continue past January 2022; \citet{Dotti2023} reported a decline of the optical flux back toward the 2018 quiescent level with no continuing oscillation, and proposed that Lense--Thirring precession of an accretion disk around a single SMBH may explain the original flare. \citet{Masterson2023} carried out high-cadence \nicer\ monitoring through August 2022 and found no periodic or quasi-periodic modulation in the X-ray, UV, or optical bands, with unusual hard X-ray flares better attributed to a variable corona than to pericenter passages. \citet{Marin2023} reported no anomalously enhanced or variable polarization with RoboPol/NOT, consistent with a stochastic active galactic nucleus (AGN) red-noise origin \citep{Vaughan2016}. \citet{Hoshi2024} found that the broad \halpha\ profile evolves on continuum-driven (light-travel) timescales rather than on orbital timescales, weakening the binary interpretation of the line variability. Multi-epoch VLBI observations \citep{An2022,An2026} resolve a single compact core at sub-milliarcsecond scales with no evidence for a binary morphology; \citet{An2026} interpret a delayed radio flare as outflow- or jet-base-driven rather than merger-driven. The cumulative weight of this evidence is that the imminent-merger interpretation is now disfavored, though not strictly refuted.

\subsection{Independent UV tests with \hst}

Two independent diagnostics of the binary scenario were not addressed by the existing follow-up and motivate the present \hst\ campaign. The first is the rest-frame near-UV continuum. Some semi-analytic and hydrodynamic circumbinary disk (CBD) calculations predict a flux deficit in the near-UV portion of the spectral energy distribution (SED), arising from a central cavity carved by the secondary SMBH or from the spectral gap between minidisks bound to the individual SMBHs and the outer circumbinary disk \citep{MilosavljevicPhinney2005,Roedig2014,Farris2015,Yan2015,Guo20}.
This signature is, however, not a generic feature of all CBD simulations: general-relativistic magnetohydrodynamic calculations including minidisks, post-processed with ray tracing, find spectra that differ only modestly from single-AGN templates \citep{Roedig2014, dAscoli2018}. Related simulations \citep{Bowen2018,Combi2022,Tiwari2025} show that the existence and location of any spectral deficit depends sensitively on mass ratio, separation, eccentricity, and accretion regime. The expected feature for J22's fiducial parameters falls in the observed UV, beyond the reach of ground-based spectroscopy.

The second test is kinematic.  The 2021--2022 optical spectra of \target\ show a multi-peaked broad Balmer profile that, taken in isolation, is qualitatively similar to the line shapes produced by relativistic accretion disks at the outer radii of single AGN, the so-called ``disk-emitter'' class \citep{EracleousHalpern1994,Eracleous2003,Eracleous2009}. Distinguishing a disk-emitter from a true binary on Balmer profiles alone is generally not possible. The critical lever arm is the comparison with \mgii. In the standard disk-emitter model, the low-ionization \mgii\ line is produced by the same dense, externally-illuminated rotating disk as the Balmer lines and should display a comparable double-peaked profile, as observed in the prototypical disk-emitter Arp\,102B with \hst/FOS UV spectroscopy \citep{Halpern1996} and across the broader \hst\ disk-emitter sample \citep{Eracleous2004,Eracleous2009}. In contrast, if the broad-line emission arises from a single bound broad-line region (BLR) around one SMBH in a binary, \mgii\ and the broad Balmer lines should be similarly profiled and share a coherent velocity centroid offset from the systemic velocity of the \oiii\ narrow-line. This test applies to binaries whose separation exceeds the BLR size ($\sim0.01-0.1$ pc). For the late-inspiral configuration of J22, the orbital separation ($\sim10^{2}\ R_{\rm g}$) is far smaller than the BLR ($\sim10^{3}-10^{4}\ R_{\rm g}$), so the BLR would be circumbinary and no per-black-hole centroid offset is expected. At J22's separation, our kinematic test therefore mainly constrains the disk-emitter alternative. This relies on the well-established reverberation-mapping result that \mgii\ samples broad-line gas at radii similar to those of \hbeta\ \citep{Homayouni2020}, in contrast to \civ, which originates at smaller radii and is dominated by a radiation-driven accretion-disk wind.

We emphasize a corollary: \civ\ is not a useful centroid anchor for binary radial-velocity tests given the general lack of a correlation between \civ\ and the Balmer lines and profile asymmetry suggesting non-virialized motion \citep[e.g.,][]{Vietri20}. Broad \civ\ is ubiquitously blueshifted relative to systemic in normal AGN by hundreds to thousands of km s$^{-1}$ \citep{Richards2002,Richards2011,Coatman2017,Temple2021}, and the disk-emitter literature has long noted that \civ\ in real disk-emitters is single-peaked because it arises in the high-ionization wind component rather than in the disk \citep{Halpern1996,Eracleous2009}. In what follows we therefore treat \mgii\ as the primary low-ionization UV diagnostic and report
the \civ\ profile separately as a wind constraint.

We obtained a 7-orbit \hst\ program (GO-16926; PI Liu) to perform these tests. Section~\ref{sec:obs} summarizes the observations and reduction. Section~\ref{sec:uv} presents the UV continuum analysis and the deficit (``notch'') search. Section~\ref{sec:lines} presents the \mgii\ versus Balmer kinematic test and the \civ\ wind constraint. Section~\ref{sec:host} presents the host-galaxy decomposition.  Section~\ref{sec:discussion} discusses our results in the context of the post-2022 follow-up record. We adopt a flat $\Lambda$CDM cosmology with $H_0 = 70\kms\,\mathrm{Mpc}^{-1}$, $\Omega_m = 0.3$.

%-----------------------------------------------------------------
\section{Target and Observations}\label{sec:obs}

\subsection{Target Properties}

\target\ is a Seyfert\,1 galaxy at spectroscopic redshift $z=0.0811$, hosted in a red, bulge-dominated elliptical or S0-like galaxy \citep{Jiang2022arxiv}. The original SDSS spectrum (2005) shows narrow emission line ratios typical of a Seyfert\,1/intermediate-type AGN and a moderately blueshifted broad \halpha\ component with offset velocity $\sim 2400\kms$. Correlated optical, near-UV, X-ray, and mid-infrared variability began around 2016 in archival difference imaging. The mid-IR flare in NEOWISE  is consistent with delayed dust reprocessing \citep{Jiang2021}.

\begin{figure*}
\centering
\includegraphics[width=\textwidth]{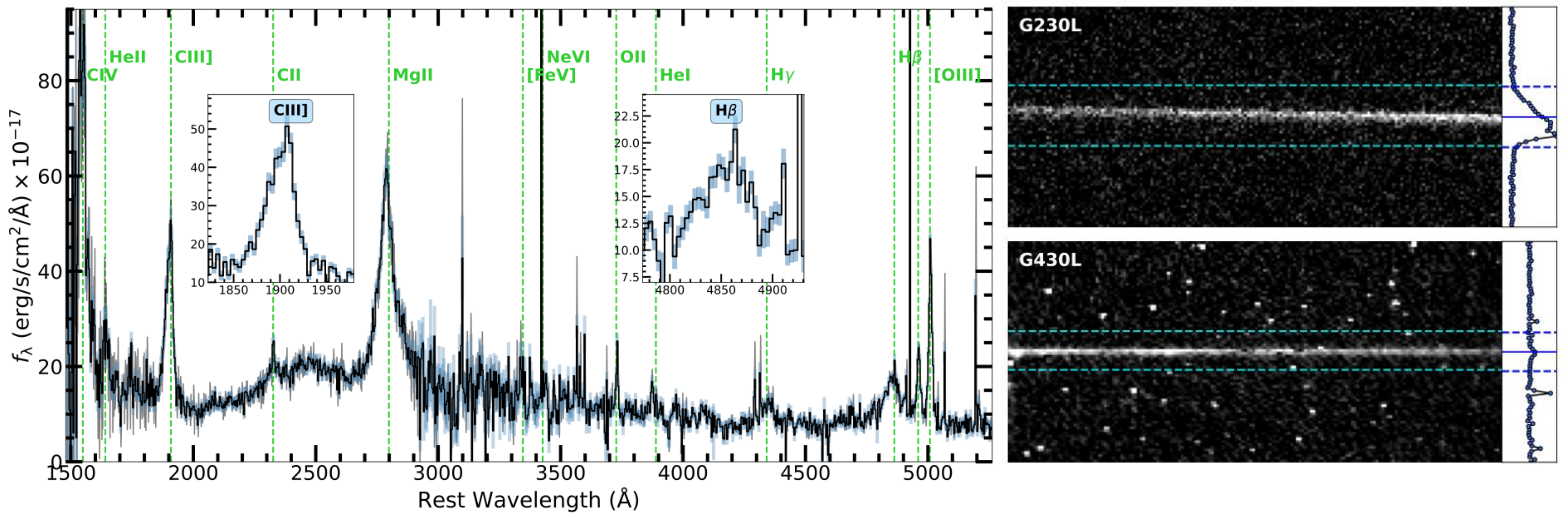}
\caption{{\bf Left:} HST/STIS G230L+G430L coadded spectrum of \target. As a check, the spectrum has been shifted to the rest frame based on the known SDSS redshift. We bin the spectrum by a factor of 2 (black) over the native resolution spectrum (gray) for readability. Uncertainties are shown in light blue. The inset panels are zoom ins of prominent line complexes. Typical quasar emission lines are marked with vertical green dashed lines and labeled. {\bf Right:} Examples of the spectral extraction for both STIS gratings. The extraction regions are indicated by dashed lines in both the 2D images and 1D profiles.  } \label{fig:coadd}
\end{figure*}
\subsection{HST/STIS UV Spectroscopy}

We obtained \hst/STIS spectroscopy on 18 August 2022 using two configurations. NUV-MAMA spectra were taken with the G230L grating (3 orbits, total exposure $\sim$\,7800\,s) covering the observed-frame range 1570--3180\,\AA\ at resolving power $R\sim 500$. The CCD spectrum was taken with the G430L grating (1 orbit, total exposure $\sim$\,1900\,s) covering 2900--5700\,\AA\ at $R\sim 530$. We used a $52''\times0\farcs2$ slit. The G230L MAMA configuration was preferred over the CCD G230LB grating because G230LB does not extend blueward of $\sim$1680 \AA\ and would miss \civ\ (observed at $\sim$1675 \AA), and because the MAMA offers higher throughput at the blue end. The two gratings overlap
over 2900--3180\,\AA, providing a control region to test for
systematic effects from intrinsic variability between visits.

Reduction was performed with the standard \texttt{calstis} pipeline, with hot-pixel masking on the MAMA data. The wavelength solution was verified using contemporaneous wavecal exposures. One-dimensional spectra were extracted with a $0\farcs2$ aperture using \texttt{x1d}, appropriate for an unresolved nucleus given the $\sim 0\farcs1$ STIS spatial resolution. In the $\sim$280 \AA\ G230L--G430L overlap region (2900–3180 \AA),the two spectra agree to within $\sim 5\%$ after standard flux calibration, indicating consistent relative flux calibration between the two gratings. We coadd all exposures of both gratings simultaneously using the HASP \texttt{swrapper} routine. At this stage, we do not shift the spectra to the rest frame using the systemic redshift from the narrow \oiii\ line, since this is incorporated natively within our spectral fitting framework. The coadded spectrum and extraction regions used in the reduction are shown in Fig.~\ref{fig:coadd}.

\begin{figure}
\centering
 \includegraphics[width=\columnwidth]{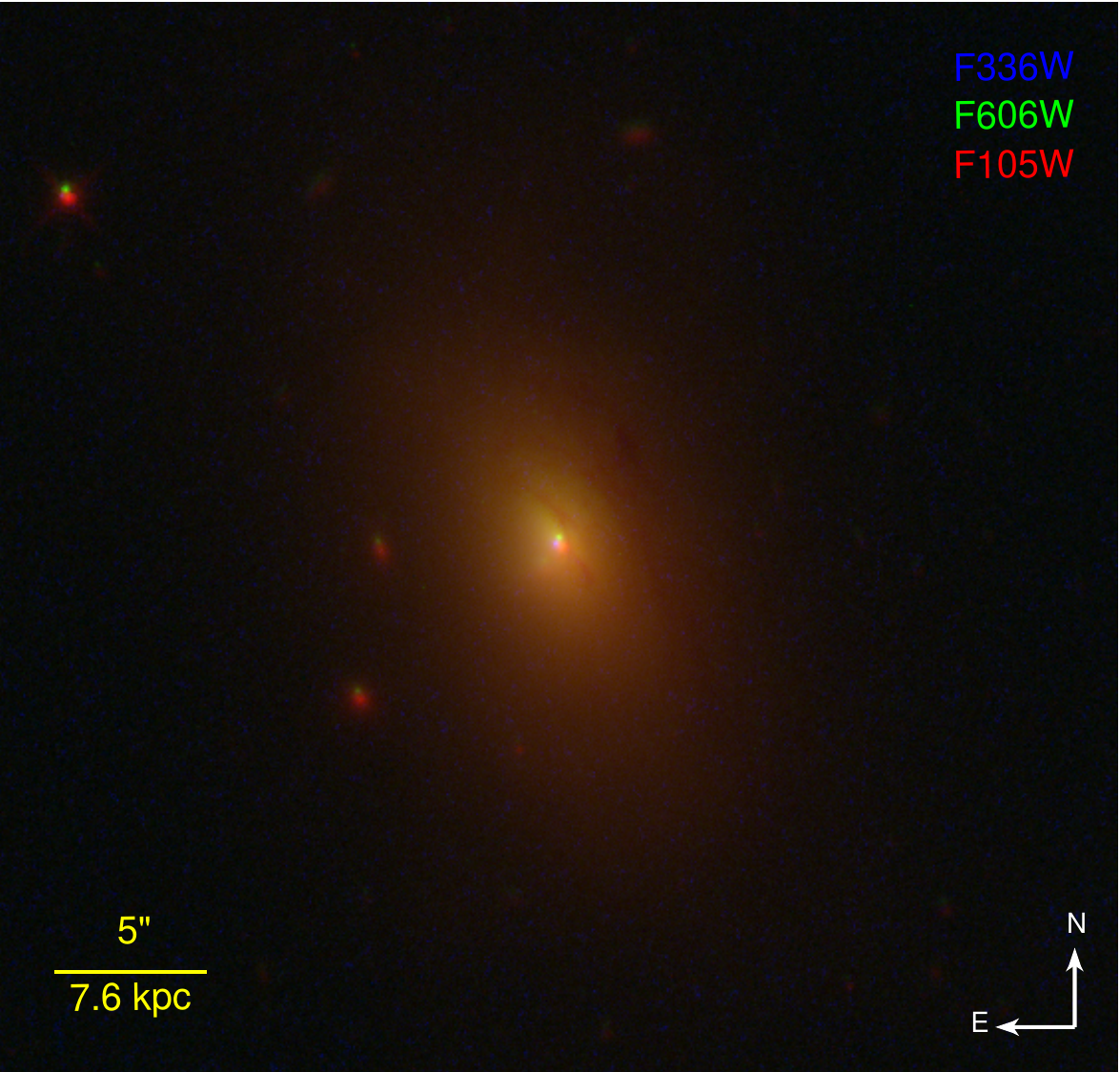}
\caption{HST/WFC3 composite image of the \target, stacking F336W (blue), F606W (green), F105W (red), each with a logarithmic stretch to highlight low-surface-brightness structure. The orientation is north up, east left.\label{fig:color_imaging}}
\end{figure}
%-----------------------------------------------------------------

\subsection{HST/WFC3 Imaging}

We obtained \hst/WFC3 three-band imaging on 7 December 2022 in UVIS/F336W, UVIS/F606W, and 15 June 2022 in IR/F105W (1 orbit each). Full detector arrays were used to provide field of view for empirical PSF stars and absolute astrometry, with an 8-point dither pattern for cosmic-ray and bad-pixel rejection and PSF oversampling. Total science exposure times were $\sim$\,2100\,s in each filter. At z $=$0.081 the filters sample rest-frame $\sim$3100 \AA\ (F336W, sensitive to recent star formation), $\sim$5450 \AA\ (F606W, roughly V band), and $\sim$9760 \AA\ (F105W, tracing the older stellar population).

UVIS data were reduced with \texttt{calwf3} with default flags and combined with \texttt{drizzlepac} on a $0\farcs04\,\mathrm{pix}^{-1}$ grid; IR data were combined on a $0\farcs06\,\mathrm{pix}^{-1}$ grid. The three frames show a slight astrometric offset seen in Fig.~ \ref{fig:color_imaging}; all three drizzled images share a common WCS solution based on Gaia reference frame. We attempted to find better astrometric alignment by reprocessing with the \texttt{tweakreg} routine on the constituent flt frames, but our exposures did not have sufficient stars to converge on a new WCS solution. For the purpose of our subsequent analysis, the lingering offset does not prevent us from drawing conclusions regarding number of model components since each filter will be fitted separately, but we note that there will be some coordinate differences on the scale of several pixels.

Empirical PSFs for each filter were constructed via \texttt{photutils} from bright, isolated stars in the WFC3 field of view for the F606W and F105W images. F3336W had no field stars within the FOV, so we construct a PSF from archival datasets (OBSIDs) sharing the same setup and observed close in time ($\sim1$ month) to our science frame. Because we have re-drizzled our images, and to better account for the HST ``breathing" effect, we use these empirically extracted PSF models rather than the focus-dependent STDPSF models provided by STScI \citep{Lallo05}. We tested the STDPSFs, but they produced significantly worse fits ($\Delta\chi^{2}/\nu\gtrsim6$).

\section{The UV Continuum and Notch Search}\label{sec:uv}

\begin{figure*}
\centering
 \includegraphics[width=\textwidth]{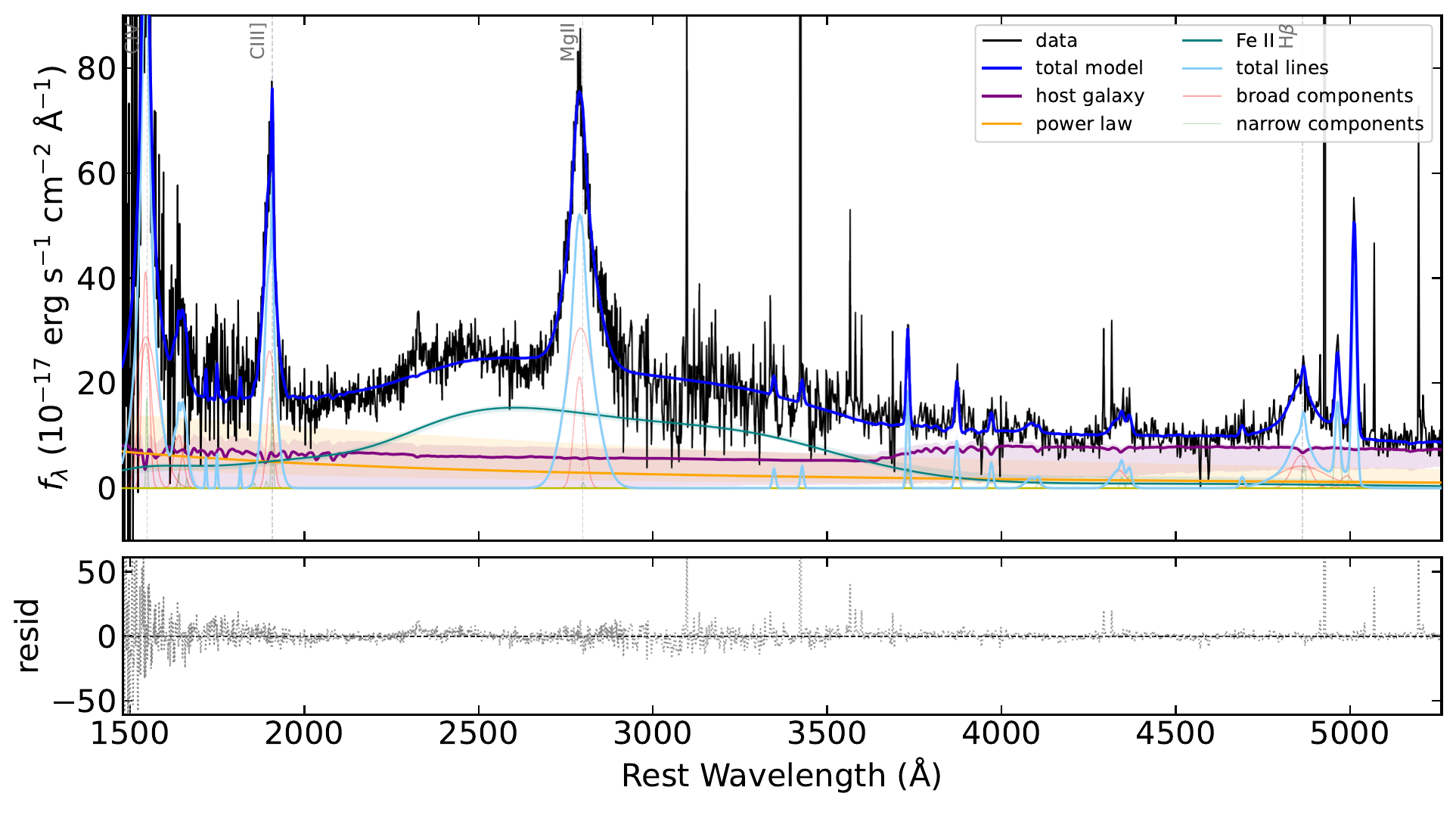}
\caption{Best-fit of the UV-optical spectrum from \hst /STIS. The lower panel shows the residuals of the fit; between 3000 and 3500\AA\, and within the \hbeta\ complex, the unfitted spikes are due to cosmic ray artifacts that could not be effectively cleaned out. The broad line complexes are best decomposed with multiple broad components. Components are color-coded based on the legend. The host galaxy and power law components contribute roughly equally to the UV end of the spectrum (although there is some degeneracy here), whereas the Fe {\sc ii} continuum dominates the NUV end. The Balmer continuum is also fitted, but has a very minor contribution in the best fit. Shading shows 1-$\sigma$ uncertainty in a given component over the posterior draws.  \label{fig:qsofit}}
\end{figure*}

\subsection{UV+Optical Spectral Fitting}

We fit the coadded STIS spectrum using \texttt{JAXQSOFIT} via the overall package \texttt{JAXSEDFIT} \citep{jaxqsofit, jaxsedfit}. The spectral fitting procedure is based on the \texttt{PyQSOFit} architecture of \citet{Shen19}, but with a key difference in handling the host galaxy and AGN continuum and lines simultaneously, while also sampling the posterior distributions via \texttt{Numpyro}. This propagates parameter degeneracies into the uncertainties when decomposing decomposing noisy spectra that likely have a complex mix of AGN and host properties, as suspected for our \target\ spectrum. 

We run the spectral fitting using the staged initialization and full posterior sampling of {\texttt OPTAX} and {\texttt NUTS}. Galactic foreground extinction ($E(B-V)=0.027$\,mag) is corrected following \citet{Cardelli1989} with $R_V=3.1$, handled via the the \texttt{dustmaps} \citep{Green18}. We use a flexible star formation history (SFH) model for the host galaxy using the continuum-only differentiable stellar population synthesis templates via \texttt{dsps} \citep{Hearin23}. Using a weighted mixture of templates is preferred here since the delayed SFH model did not lead to a convergent fit, and the host is a bulge-dominated, S0-like galaxy (\S~\ref{sec:host}). We apply the instrumental resolution correction to forward model the line spread function of STIS resolution broadening and recover the intrinsic emission line widths on all posterior samples. 

We fit the UV+optical continuum with power-law accretion-disk continuum plus the optical-UV ``small blue bump'' from \ion{Fe}{2} blends and Balmer continuum, parameterized following standard quasar SED prescriptions \citep{VandenBerk2001}. The priors on fitted emission lines are imported from the \texttt{JAXQSOFIT} default line priors table, which we supplement with the emission line galaxy table and high-ionization/coronal line table to cover several suspected emission lines. While it would likely yield a marginal improvement to the fit statistics, we do not mask the $\sim$3 uncleaned cosmic ray artifacts during the fit so as not to mask out nearby low-level emission lines; our fits do not mistakenly yield any narrow line components at the locations of the artifacts. To sample the posterior distribution and obtain meaningful uncertainties, we run a warm-up of 1000 steps followed by 2000 samples in each of 4 chains. All reported parameters have ${\rm \hat R}<$1.01 and bulk and tail effective sample sizes above 400. 

\begin{figure*}
\centering
 \includegraphics[width=\textwidth]{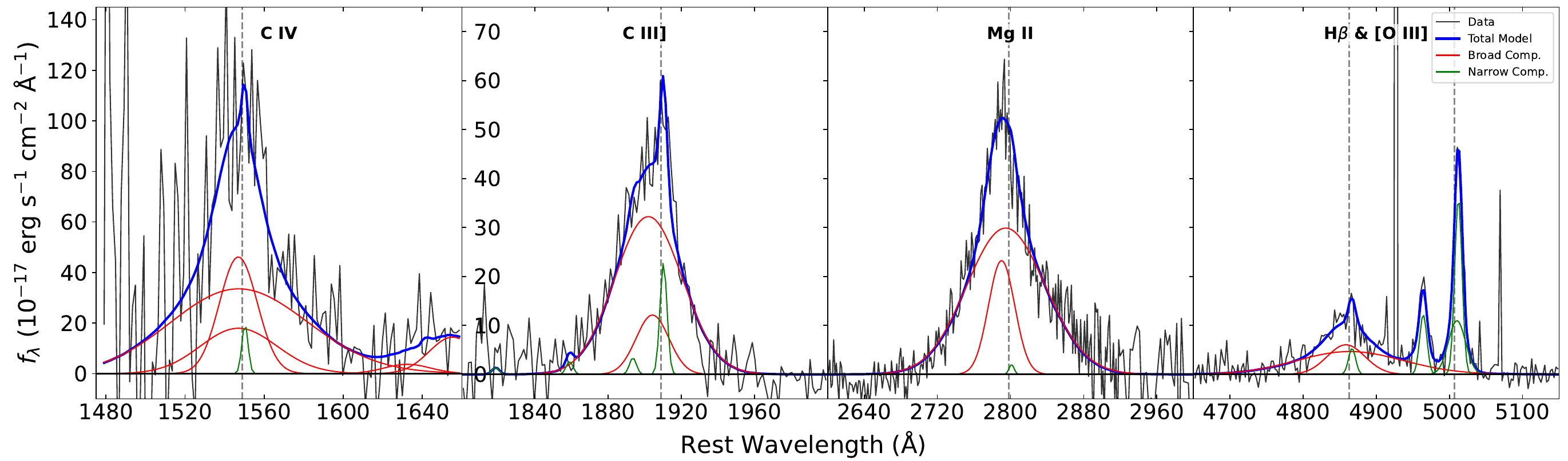}
\caption{Zoom-ins of emission line regions from Fig \ref{fig:qsofit}. The data and models are continuum-subtracted. Components are color-coded based on the legend. Vertical dashed lines indicate the rest-frame wavelength of the named emission lines. The three right-most panels all share a common y-axis. \label{fig:qsofit_sub}}
\end{figure*}

The resulting full spectral model is shown in Fig.~\ref{fig:qsofit}. Zoom-ins on the decomposition of key quasar emission line complexes are shown in Fig.~\ref{fig:qsofit_sub}. We list the properties of the relevant broad lines in Table~\ref{tab:emi_lines}. All other emission line fits are reported in Appendix Table~\ref{tab:emi_lines_appendix}.

The robustness of the best-fit model can be quantified through several fit statistics: we obtain a $\chi^{2}$ per pixel = 7.515, although more informative here is the weighted root mean square, wrms = 2.741. The $\chi^{2}$ value excludes the fitted fractional jitter (posterior $\sim6\%$), and is thus less reflective of the actual goodness of fit. Several under-fitted regions partially drive the fit statistic. This is most apparent in several regions of the spectrum: at the edges of the spectral coverage of both gratings ($\sim$1500 \AA\ and $\sim$3200 \AA\ for G230L and G430L, respectively) there is considerable noise. The posterior probability distributions for the best-fit are shown in Fig.~\ref{fig:qsofit_corner}. Here too, it is apparent that the complex modeling is limited by parameter degeneracies, e.g., the non-gaussian multi-peaked profile between the continuum fraction apportioned between the power law and host. The power law, $f_{\lambda}\propto \lambda^{\alpha}$, yields a best-fit slope of $\alpha=-1.78\pm$0.04 with normalization 2.72$^{+7.68}_{-1.38}\times10^{-17}{\rm erg\ s^{-1}\ cm^{-2}\AA^{-1}}$ at $\lambda = 5100\AA$; this is slightly flatter than a standard thin disk \citep{Shakura73}, either due to a combination of reddening and contamination from the host galaxy, or degeneracy in the model fitting. We also note that the fitted stellar velocity dispersion, $\sigma_{*} \approx 111^{+71}_{-29}\kms\ $is well below the STIS spectral resolution of (FWHM$\sim600\kms$, i.e., $\sigma_{\rm instr}\sim250\kms$), so we do not further consider this in our interpretation. 

We note that the broadened \ion{Fe}{2} continuum clearly under-fits the complex structure in the UV; in trials using only the Optax inference method, we achieved fits including a higher spectral resolution \ion{Fe}{2} template match particularly around 2500\AA, which yielded $\chi^{2}/pix \lesssim$6 and wrms $\lesssim$ 2. We opt instead for full posterior sampling so that the uncertainties on the individual line components can be robustly characterized. However, in the well sampled model, the \ion{Fe}{2} component does not exhibit much scatter across samples, as compared to the degeneracy indicated by large shaded uncertainty regions for the nearly comparable power law and host components. We note that the emission line properties appear to be fairly robust such that the more resolved \ion{Fe}{2} does not meaningfully change our interpretation of the line properties. The posterior Fe {\sc ii} broadening (FWHM $\approx 5.5 \times 10^{4} \kms$) is unphysically large, so the template effectively acts as a smooth pseudo-continuum. In Optax fits with the Fe {\sc ii} width bounded to $\lesssim$ 10,000 $\kms$, the broad Mg {\sc ii} and \hbeta\ centroids and widths change by less than $\lesssim 200 \kms$ and $\kms$$\lesssim 370$ $\kms$, respectively, within their 1$\sigma$ uncertainties.

\subsection{UV Continuum Notch Search}

A phenomenological notch or cutoff, modeled as a Gaussian absorption profile in $\log F_\nu$, motivated by the CBD predictions of \citet{Roedig2014, Farris2015, Yan2015, Guo20}, is a potential observable signature of a physical gap between a SMBH binary CBD and the interior minidisks. However, the single-AGN model leaves no coherent residual deficit that would match such a notch. Furthermore, the \ion{Fe}{2} template is degenerate with such a notch absorption feature, and while the template visibly under-fits the spectrum around 2500 \AA\ where a notch might most plausibly be centered, the systematics of the fit prevent us from placing a robust upper limit on a non-detection. 

The non-detection is informative but, on its own, not decisive against the binary scenario. A near-UV deficit is a possible, not a universal, prediction of CBD models. The GRMHD simulations of \citet{dAscoli2018}, post-processed with ray tracing, for example, produce SEDs that differ only modestly from those of single AGN: the inclusion of minidisks bound to the individual SMBHs fills in the spectral region in which a notch would otherwise appear. \citet{Roedig2014} attribute the absence of a notch in those calculations to the comparative faintness of the minidisks at the binary separation analyzed. Related simulations \citep{Bowen2018,Combi2022,Tiwari2025} find that the existence and depth of any deficit depends on mass ratio, separation, eccentricity, and accretion regime. The fiducial CBD location predicted by J22 also depends on the assumed inner-edge radius and accretion mode \citep{Guo20}; our qualitative non-detection therefore cannot exclude CBD configurations that predict shallow deficits or deficits outside our band, and does not refute the binary scenario as a class.

%-----------------------------------------------------------------
\section{Broad Emission Line Profiles}\label{sec:lines}

\subsection{The Balmer-line context}

The \hbeta\ and \halpha\ profiles of \target\ have evolved dramatically over the past two decades \citep{Jiang2022arxiv,Hoshi2024, He26}. The 2005 SDSS spectrum shows a single broad \halpha\ component blueshifted by $\sim 2400\kms$ from the narrow-line systemic. The post-2021 P200/DBSP and LJT spectra are well modeled by three Gaussians: a near-systemic component, a strongly blueshifted component at $\sim -4000\kms$, and a strongly redshifted component at $\sim +4600\kms$. Such extreme velocity offsets and double-peaked profiles are also found in disk-emitters \citep{EracleousHalpern1994,Eracleous2003,Eracleous2009}; distinguishing the two scenarios on Balmer profiles alone is not possible. As a kinematic check, \citet{Hoshi2024} reported that the post-flare \halpha\ profile evolves on continuum-driven (light-travel) timescales rather than on orbital ones, which is in tension with a binary origin for the line variability and which we revisit in \S~\ref{sec:discussion}.

\subsection{The diagnostic: Mg II versus Balmer}

Our primary kinematic test is the comparison of \mgii\ with the broad Balmer profile. This test exploits the empirical fact that, in prototypical disk-emitters such as Arp\,102B, \mgii\ has nearly the same profile as the Balmer lines, with peaks separated by $\sim 12{,}000\kms$ and a typical Mg\,{\sc ii}/H$\beta$ ratio near unity \citep{Halpern1996}. In the broader \hst\ disk-emitter sample \citep{Eracleous2004,Eracleous2009}, \mgii\ is consistently the only UV resonance line that mirrors the Balmer double-peak structure. This is a consequence of the dense, externally-illuminated outer-disk gas producing both \mgii\ and the Balmer lines, while the high-ionization wind dominates Lyman$\alpha$ and \civ. The reverberation-mapping result that \mgii\ traces broad-line gas at radii comparable to \hbeta\ \citep{Homayouni2020} provides the physical basis for this expectation in any AGN, single or binary.

\begin{figure}
\centering
 \includegraphics[width=0.49\textwidth]{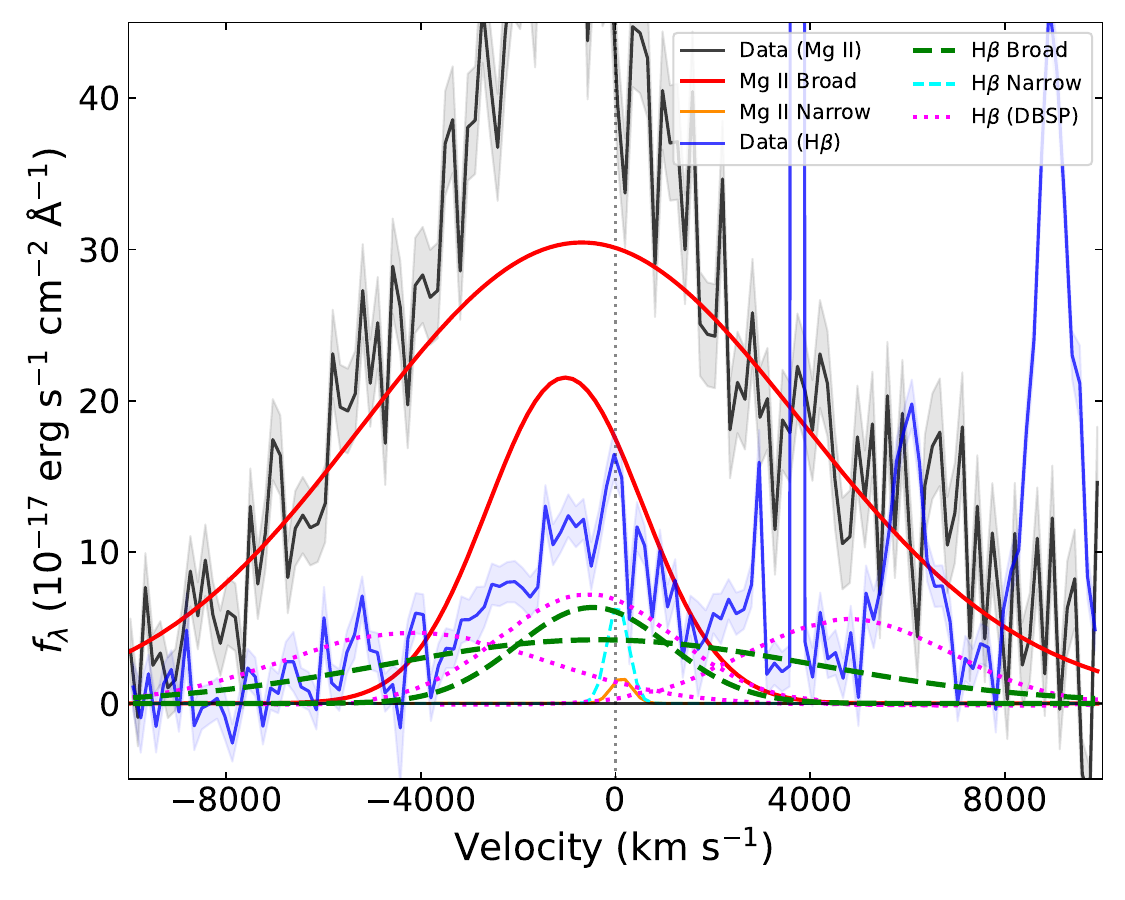}
\caption{Broad emission line profiles in velocity space. \mgii\ (black) and \hbeta\ (blue) from the new STIS spectrum, normalized to the local continuum, as well as the broad and narrow components of each. We also plot the best-fit broad components for \hbeta\ from the post-flare DBSP spectrum \citep{Jiang2022arxiv} (pink).  The vertical dashed line marks the systemic redshift from \oiii. This is the kinematic test of disk-emitter versus single bound BLR. The narrow \mgii\ and \hbeta components are tied in velocity and width in the fit, so their agreement is by construction. Both broad \mgii\ components are blueshifted relative to their \hbeta\ counterparts (\S 4.2). 
\label{fig:line_profiles}}
\end{figure}

We measure \mgii\ in the STIS spectrum, with the local continuum modeled as a power law plus a phenomenological \ion{Fe}{2} pseudo-continuum template \citep{VestergaardWilkes2001}. One of the key advantages of \texttt{JAXQSOFit} over previous algorithms is that the emission lines are modeled simultaneously with the continuum fit, instead of {\emph {ex post facto}}, which should yield more robust emission line fits. Our findings are summarized in Figure~\ref{fig:line_profiles}: 
\begin{enumerate}
    \item \mgii\ is well described by a single-peaked profile, where a combination of two broad components and a single narrow component provides the best fit.
    \item The velocity centroid of the principal broad \mgii\ component is offset from the \oiii\ systemic velocity by -1111$^{+133}_{-103} \kms$. The secondary broad component is offset by -776 $^{+183}_{-110} \kms$.
    \item The corresponding centroid offset of the principal broad \hbeta\ component is -467$^{+135}_{-122}\kms$, and that of the secondary is -336$^{+253}_{-149}\kms$. This is comparable to the offset of the main \hbeta\ broad component in the post-flare DBSP spectrum of J22 (-412 $\kms$). However, our best-fit model only requires two broad components for \hbeta\ compared to the three components from J22. The velocity offset of the secondary broad component in our fit is broadly consistent with the primary component within the fitting uncertainties. It may be that the symmetrically shifted additional peaks found in the DBSP spectrum have faded to near pre-flare levels such that only a relatively minor, but substantially broadened (FWHM $\sim10500$ km/s), secondary broad component is distinguishable against the more narrow primary broad component, masking the true profile containing the twin peaks (see Fig.~\ref{fig:line_profiles}); however, differentiating these scenarios is not possible at the resolution of the current spectrum.
    \item The line widths (FWHM) are $3756^{+318}_{-294}$$\kms$ and $10643^{+808}_{-616}$$\kms$ for the two broad components of \mgii\ and $3676^{+378}_{-256}$$\kms$ and $10458^{+1275}_{-1895}$$\kms$ for \hbeta. The primary (and secondary) broad components of \mgii\ and \hbeta\ agree well within their 1-$\sigma$ uncertainties.
\end{enumerate}

\begin{deluxetable*}{lcccr}
%\tabletypesize{\script} \tablewidth{8pt}
\tabletypesize{\normalsize}
\tablewidth{\textwidth}

\tablecaption{Emission Line Properties\
\label{tab:emi_lines}}
\tablehead{ 
\colhead{Emission} & \colhead{Luminosity}& \colhead{Velocity Offset} & \colhead{FWHM} & \colhead{EW} \\
\colhead{Line} & \colhead{log[erg s$^{-1}$]} & \colhead{km s$^{-1}$} & \colhead{km s$^{-1}$} & \colhead{\AA} 
}
\startdata 
\textbf{C \sc{iv}} & $41.19^{+0.16}_{-0.15}$ & $-471^{+221}_{-250}$ & $4111^{+587}_{-428}$ & $55.4^{+25.2}_{-16.3}$ \\
1549.06\AA & $41.30^{+0.15}_{-0.19}$ & $-407^{+155}_{-181}$ & $8135^{+1806}_{-709}$ & $70.4^{+27.9}_{-25.4}$ \\
 & $41.61^{+0.07}_{-0.09}$ & $-359^{+295}_{-116}$ & $16336^{+2201}_{-1398}$ & $143.8^{+23.3}_{-27.6}$ \\
\noalign{\smallskip}
\textbf{C \sc{iii}]} & $40.85^{+0.51}_{-0.35}$ & $-947^{+251}_{-125}$ & $3705^{+1771}_{-692}$ & $25.4^{+57.7}_{-13.9}$ \\
1908.73\AA & $41.33^{+0.07}_{-0.60}$ & $-1005^{+198}_{-127}$ & $7464^{+2377}_{-684}$ & $77.7^{+13.7}_{-58.3}$ \\
\noalign{\smallskip}
\textbf{Mg \sc{ii}} & $41.11^{+0.11}_{-0.05}$ & $-882^{+132}_{-102}$ & $3756^{+318}_{-294}$ & $33.9^{+9.7}_{-3.6}$ \\
2798.75\AA & $41.72^{+0.01}_{-0.03}$ & $-548^{+182}_{-109}$ & $10643^{+808}_{-616}$ & $139.4^{+4.6}_{-10.3}$ \\
\noalign{\smallskip}
\textbf{H$\beta$} & $40.82^{+0.08}_{-0.10}$ & $-239^{+134}_{-120}$ & $3676^{+378}_{-256}$ & $47.1^{+9.4}_{-9.7}$ \\
4862.68\AA & $41.09^{+0.05}_{-0.06}$ & $-108^{+252}_{-148}$ & $10458^{+1275}_{-1895}$ & $88.8^{+9.9}_{-12.1}$ 
\enddata
\tablecomments{ 
Emission line properties measured from the spectral fitting of the coadded $HST$/STIS and NUV-MAMA spectra in rest-frame. Luminosity is reported in log units of erg s$^{-1}$, while the velocity offset from the vacuum central wavelength and the full width at half max (FWHM) are in units of km s$^{-1}$. All values are for the broad components of the line, and are quoted as the parameter's median value across draws. Errors are quoted at the 1-$\sigma$ level from Markov Chain Monte Carlo simulations. Multiple broad components for a line are listed sequentially. Velocity offsets are relative to the vacuum wavelengths at z = 0.0811; the \oiii\ core lies at $+228^{+18}_{-15}$$\kms$ on this scale, so offsets relative to [O {\sc iii}] are $v - 228 \kms$.
}
\end{deluxetable*}

Because Mg {\sc ii} is measured in G230L and the [O {\sc iii}] systemic in G430L, their relative velocity carries a zero-point uncertainty from target centering in the 0\farcs2 slit. A 0\farcs02 offset shifts the G230L and G430L wavelength scales by $\sim$125 and $\sim$60 $\kms$, respectively, so we adopt $\sim$140 $\kms$ as the systematic uncertainty on cross-grating velocity differences. The narrow Mg {\sc ii} and \hbeta\ components are tied in our fit, so their agreement does not test this zero-point. In comparing between \mgii\ and the broad
Balmer profile, two regimes are distinguishable:
\begin{itemize}
    \item \emph{Disk-emitter prediction.}  \mgii\ is double-peaked with peak structure that mirrors that of \hbeta/\halpha, as in Arp\,102B.
    \item \emph{Single bound BLR around one SMBH (binary or recoiling SMBH).}  \mgii\ and the broad Balmer lines are similarly profiled (no relativistic disk double peak) and share a coherent velocity centroid offset from \oiii\ systemic.
\end{itemize}
At the HST epoch (2022 August), both \mgii\ and \hbeta\ are single-peaked and well described by a core (FWHM$\approx3600-3800 \kms$) plus wing (FWHM$\approx10400-10700 \kms$) pair of Gaussians whose widths agree within 1$\sigma$. Neither line shows the displaced twin peaks of a relativistic disk-emitter, and the shoulders at $\sim$-4000 and $\sim$+4600 $\kms$ in the 2021–2022 Balmer spectra of J22 are absent. The dominant peak in both lines is towards the red end with the shoulder towards the blue end, inconsistent with the typical profile for a Keplerian disk with Doppler boosting; however, such profiles can be created in disk-emitters with non-axisymmetric disks. The broad \hbeta\ components lie within $\sim 500 \kms$ of [O {\sc iii}], while the broad \mgii\ components are blueshifted by $\sim 800-1100 \kms$. The Mg II–\hbeta\ offsets are -643 $\pm 222 \kms$ for the cores and -440 $\pm 284 \kms$ for the wings, including the cross-grating systematic ($\sim2.9\sigma$ and $\sim1.5\sigma$, respectively). Broad Mg {\sc ii} and \hbeta\ velocity shifts of a few hundred $\kms$ occur in normal quasars \citep{Shen16}, and Fe {\sc ii} template systematics (\S3.1) mainly affect the Mg {\sc ii} wing. We therefore do not regard this offset as evidence for a distinct kinematic component, and thus find no evidence for a disk-emitter at this epoch, nor for the large coherent centroid offset expected from a BLR bound to one SMBH. We caution that this test, while informative against the classical disk-emitter scenario, does {\emph not} uniquely indicate a binary: a recoiling SMBH carrying a bound BLR \citep{ShenLoeb2010,Eracleous2012} produces the same kinematic signature, as does a transient outflow disturbing a single BLR \citep[e.g., for PG\,1302$-$102,][]{Rigamonti2025}. Multi-epoch monitoring is required to distinguish between these scenarios.

\subsection{The C IV profile: a wind constraint, not a centroid}

\civ\ is well-described by multiple blueshifted broad components. The main broad line has the largest blueshift of 699$^{+222}_{-251}\kms$ relative to the systemic \oiii\ velocity, with FWHM 4111$^{+587}_{428}\kms$ and equivalent width 55$^{+25}_{-16}$\AA.  These properties fall within the normal distribution for radio-quiet AGN \citep{Richards2011,Coatman2017,Temple2023}: the mean \civ\ blueshift in radio-quiet quasars is $\sim 800\kms$, with extreme values in luminous, high-Eddington-ratio sources reaching several thousand $\kms$ \citep{Coatman2017,Temple2021}. Our \civ\ measurement is therefore a constraint on a normal accretion-disk wind in \target\ rather than a kinematic test of the binary scenario. We do not use the \civ\ centroid in the disk-emitter--versus--binary comparison, in line with standard practice in the SMBH-binary radial-velocity literature \citep{Eracleous2012,Liu2014,Runnoe2017,Wang2017,Guo2019}.

To estimate the outflow rate from C {\sc iv}, we follow the prescription of \citet{Vietri20}:
\begin{equation}
\dot{M}_{\rm out} = \frac{M_{\rm out}\times v_{out}}{R_{\rm out}},
\end{equation}
where the outflowing mass from a thin spherical shell is calculated as:
\begin{equation}
M_{\rm out}\approx 100\frac{L_{\rm C IV}/10^{45} {\rm erg\ s^{-1}}}{n_{\rm e}/10^{9}\times n({\rm C^{+3}})/n({\rm C})}{M_{\odot}} .
\end{equation}
Here, $L_{\rm C IV}$ is the C {\sc iv} line luminosity, and $n_{\rm e}$ is the electron density, assumed to be 10$^{9.5}$ cm$^{-3}$. The fractional ionization of C {\sc iv}, $n({\rm C^{+3}})/n({\rm C})\simeq$ 0.5 is based on an assumed ionization parameter of $U = 0.05$ for a high-ionization cloud of $\sim 1 - 5 M_{\odot}$ metallicity. We take the outflow velocity $v_{\rm out}$ for the main blue-shifted broad component. The outflow radius $R_{\rm out}$ of \civ\ is computed from the radius-luminosity relation of \citet{Lira18}, giving $R_{\rm out} = 0.001\pm0.001$pc. This yields an outflow rate for the main \civ\ component of $\dot{M}_{\rm out} = 0.0064^{+0.0066}_{-0.0051}M_{\odot}/yr$. Summing the three \civ\ broad components yields a total outflow rate of $\dot{M}_{\rm out} = 0.0280^{+0.0174}_{-0.0128}M_{\odot}/yr$, which is on the lower end of the range found for Type-I AGN from the SUPER sample of \citet{Vietri20}. If SDSS J1430+2303 does indeed harbor a binary SMBH, the inner configuration does not appear to enhance nor inhibit the general outflow rate from the surrounding CBD.

%-----------------------------------------------------------------
\section{Host Galaxy Imaging}\label{sec:host}

\begin{deluxetable*}{lcccccccc} \tabletypesize{\small}
\tablecaption{GALFITM Results \label{tab:galfit_bestfit}}
\tablehead{ 
\colhead{Component} & \colhead{$\alpha, \delta$ (deg)} & \colhead{x} & \colhead{y} & \colhead{$m$(AB)} & \colhead{$R_{\rm e}$(pix)} & \colhead{$n$} & \colhead{Axis Ratio} & \colhead{PA(deg)} \\
\colhead{(1)} & \colhead{(2)} & \colhead{(3)} & \colhead{(4)} & \colhead{(5)} & \colhead{(6)} & \colhead{(7)} & \colhead{(8)} & \colhead{(9)}
}
\startdata
\hline
\hline
\multicolumn{9}{l}{{\bf F105W: $\chi^{2}/\nu$ = 1.364}} \\
\hline
S\'ersic &  217.566891+23.062357& 298.9$\pm$0.1 & 300.7$\pm$0.3 & 16.167$\pm$0.038 & 53.23$\pm$0.66 & 3.293$\pm$0.276 & 0.66$\pm$0.01 & 22.569$\pm$0.130 \\
S\'ersic &  217.567011+23.062143& 291.4$\pm$0.3 & 287.8$\pm$0.8 & 20.569$\pm$0.279 & 10.34$\pm$0.81 & 1.976$\pm$0.225 & 0.80$\pm$0.09 & -50.381$\pm$9.984 \\
PSF &  217.566874+23.062362& 299.8$\pm$0.3 & 301.1$\pm$0.2 & 19.298$\pm$0.253 &  &  &  &  \\
\hline
\multicolumn{9}{l}{{\bf F606W: $\chi^{2}/\nu$ = 1.213}} \\
\hline
S\'ersic &  217.566907+23.062382& 358.4$\pm$0.2 & 345.2$\pm$0.2 & 16.272$\pm$0.012 & 71.31$\pm$0.11 & 1.821$\pm$0.040 & 0.67$\pm$0.01 & 21.304$\pm$0.152 \\
S\'ersic &  217.566854+23.062424& 353.9$\pm$0.2 & 341.6$\pm$0.0 & 18.557$\pm$0.025 & 1.43$\pm$0.00 & 5.000$\pm$0.000 & 0.51$\pm$0.24 & 29.300$\pm$8.721 \\
PSF &  217.566838+23.062335& 352.4$\pm$7.3 & 349.5$\pm$6.9 & 24.13$\pm$0.807 &  &  &  &  \\
\hline
\multicolumn{9}{l}{{\bf F336W: $\chi^{2}/\nu$ = 2.157}} \\
\hline
S\'ersic & 217.566900+23.062306 & 107.2$\pm$0.2 & 102.4$\pm$2.8 & 21.979$\pm$1.398 & 17.36$\pm$9.40 & 4.656$\pm$0.196 & 0.16$\pm$0.06 & 32.211$\pm$15.862 \\
S\'ersic &  217.566995+23.062279 & 114.9$\pm$2.7 & 105.2$\pm$3.2 & 18.985$\pm$0.803 & 38.10$\pm$17.34 & 0.964$\pm$1.530 & 0.65$\pm$0.03 & 8.992$\pm$1.504 \\
PSF &  217.566890+23.062366& 106.7$\pm$0.1 & 96.7$\pm$0.1 & 18.251$\pm$0.606 &  &  &  &  \\
\enddata
\tablecomments{Best fit results for UVIS/IR image decomposition. For each of the three \hst\ filters, we list the best-fit parameters for the model that achieved the statistically best fit. Uncertainties are on each parameter are computed as the quadrature sum of the nominal error output by \texttt{galfitm} and the dispersion on that parameter found across runs using the composite PSF and each component PSF as the model reference. We list: 1) the model component type, 2) RA and DEC in decimal degrees, 3) \& 4) the pixel coordinates (see text for the imaging bounds for each filter), 5) the component's total magnitude, 6) effective radius, 7) S\'ersic index, 8) axis ratio, and 9) position angle. }
\end{deluxetable*}

We analyze the WFC3 imaging to decompose the AGN and host (bulge and disk), to search for circumnuclear star formation in F336W, and to search for low-surface-brightness tidal features in F105W. We fit the three bands independently using a consistent set of nested models. Image decomposition is performed with \texttt{AFFOGATO} \citep{Nolan25} a wrapper for \texttt{galfitm} \citep{Haussler22}. The models range in increasing complexity based on the number of S\'ersic and point-like (PSF) components. The uncertainties reported by \texttt{galfitm} are known to be underestimated; we therefore incorporate additional uncertainty on the best fit parameters based on the variance of the individual extracted PSFs that make up the overall composite PSF used for our main modeling. Each fit is run again for every constituent PSF, and the variance on a given parameter is added in quadrature to the nominal uncertainty. This source of 1-$\sigma$ uncertainty helps account for the systematics introduced by needing archival field stars to create the PSF model. 

The best fit models for each filter are given in Table \ref{tab:galfit_bestfit}. The uncertainties on some parameters are quite large in some filters. This reflects both degeneracies among the fitted parameters and the limitations of a composite PSF. We therefore do not further interpret some of the fitted S\'ersic properties beyond the bulge--disk magnitude decomposition.

For all three filters, the best-fit model contains two S\'ersic components and a single centrally located PSF component. For F606W, the 3-S\'ersic model achieves a marginally lower reduced $\chi^{2}$; however, this is likely a result of over-fitting. We compare two models using the Bayesian information criterion \citep[BIC; i.e.,][]{Claeskens08}: 
\begin{equation}
{\rm BIC}=k\times\ln(n)-2\times\ln(\mathcal{\hat{L}}), 
\end{equation}
where for nested models $\ln(\mathcal{\hat{L}}) = const\ - \chi^{2}/2$. The 3-S\'ersic model lowers $\chi^{2}$ by $\lvert\Delta\chi^{2}\rvert$ for 7 additional parameters over $n$ = [700$\times$700] pixels, giving $\Delta$BIC  $\equiv$ BIC$_{3}$ - BIC$_{2}$ = -28766. Because drizzled pixels are correlated, $n$ overstates the number of independent data points and the BIC penalty is only approximate. We therefore prefer the 2-S\'ersic model mainly on physical grounds: the third component collapses to the minimum allowed effective radius and mimics a point source.

\begin{figure*}[htbp]
    \centering
    \begin{subfigure}{\textwidth}
        \centering
        \includegraphics[width=\textwidth]{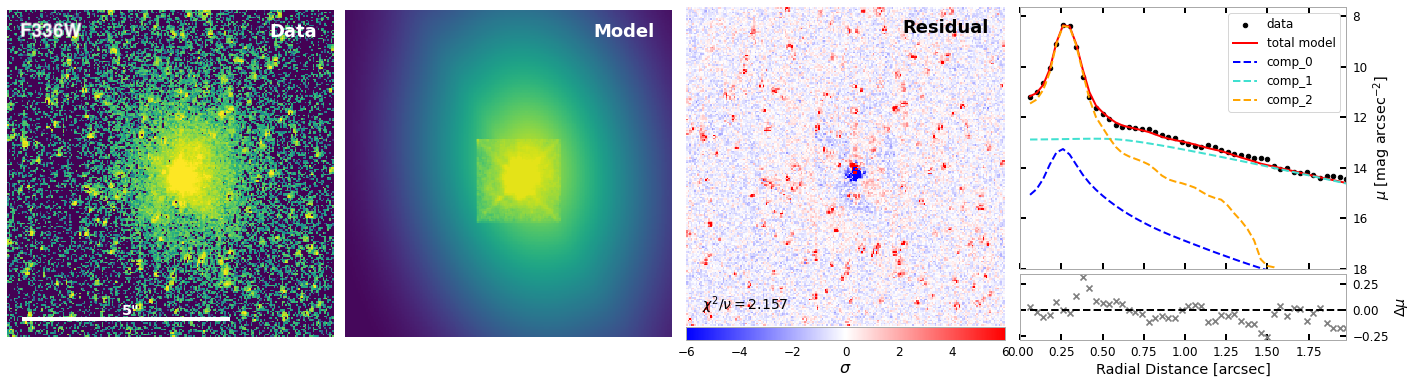}
        
        \label{fig:sub1}
    \end{subfigure}
    %\hfill % Adds horizontal spacing between the two figures
    \begin{subfigure}{\textwidth}
        \centering
        \includegraphics[width=\textwidth]{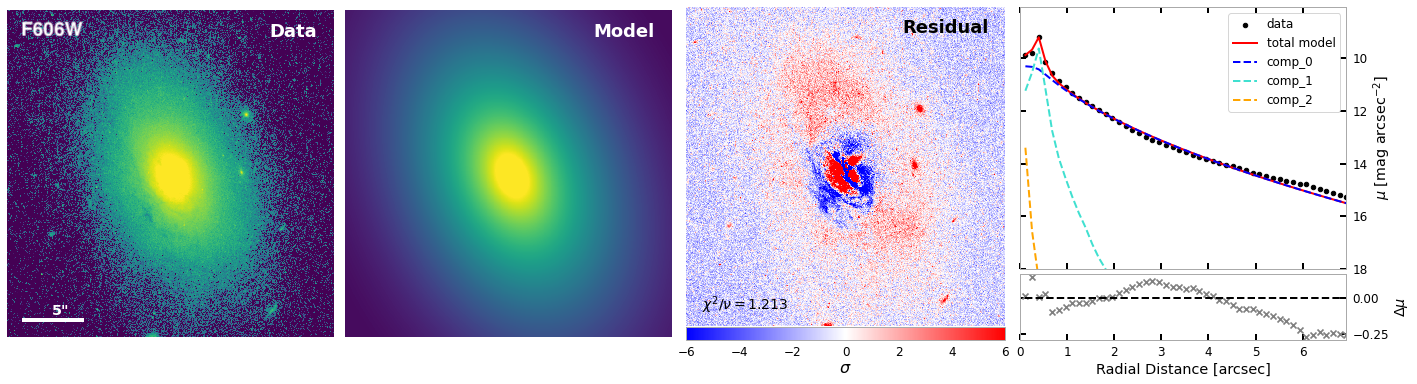}
            \label{fig:sub2}
    \end{subfigure}
    \begin{subfigure}{\textwidth}
        \centering
        \includegraphics[width=\textwidth]{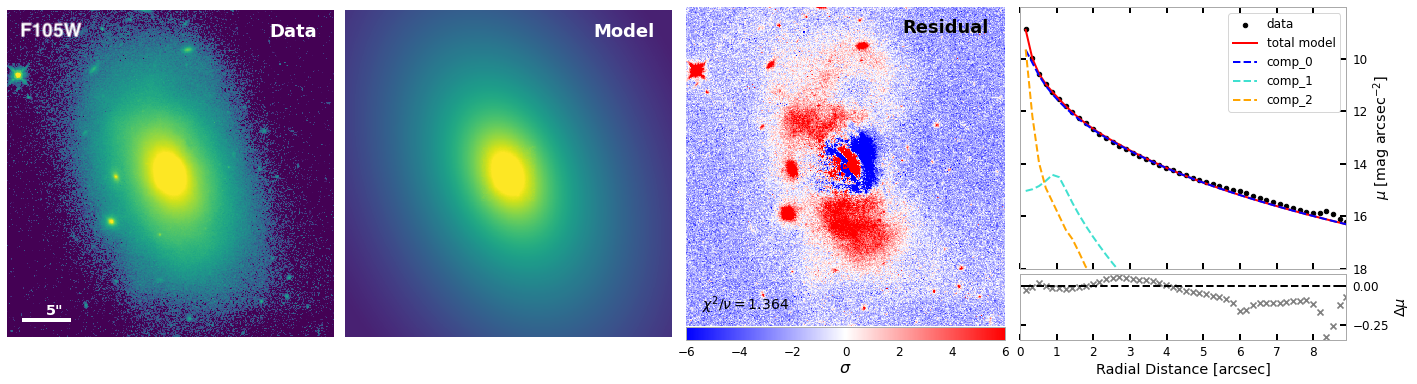}
        
        \label{fig:sub3}
    \end{subfigure}
    \caption{Best-fit image decomposition for our \hst /WFC3 imaging of the host galaxy of \target. For F336W, F606W, and F105W (from Top to Bottom), we show: the data image with a logarithmic stretch to highlight low-surface-brightness structure, the best-fit total model, the residual image scaled by significance level in the color bar, and the radial profile. The individual model components in the legend correspond to the two S\'ersic components (0 and 1) and the single PSF component (2) in all cases. The sub-panel below each radial profile shows the scaled residuals as a function of radial distance from the center of the frame. The scale bars for each filter demark 5\arcsec, although each filter covers different spatial extents due to the differing pixel sizes and observable extent of the host galaxy at that wavelength: F336W is much more centrally concentrated and finer spatial resolution, so we zoom in to focus on the central region. The fit statistic for each model is given in the lower left corner of the residual plots; In general, there are some coherent residuals left in the F606W and F105W frames; however, the foreground stars embedded within the disk's extent make masking difficult, leaving the more concentrated residuals in most cases. 
}
    \label{fig:galfit}
\end{figure*}

Even in this two S\'ersic model, one of the components shrinks down to the limit set for the effective radius thereby mimicking a PSF component. This compact S\'ersic component dominates over the PSF, leading to an unconstrained PSF mag$\sim$24 and position ($\pm7$ pix). We therefore take the nuclear flux to be the sum of these two compact components. The best-fit models are shown in Fig \ref{fig:galfit}.

The host is well resolved in all three bands and is dominated by an elliptical/S0-like light distribution centered on a bright unresolved nuclear component, consistent with the previous SDSS imaging \citep[Figure~\ref{fig:color_imaging};][]{Jiang2022arxiv}. The two S\'ersic components make up a typical bulge plus disk morphology. No tidal arms or shells are detected to a 3$\sigma$ surface-brightness limit of $\mu$(F105W) $\approx$ 25.06 mag arcsec$^{-2}$ (measured over $\sim$67 arcsec$^{-2}$). This does not rule out a past major merger, since tidal features fade on Gyr timescales. 

There are varying amounts of residuals in each filter. Our F336W models leaves very few residuals of statistical significance aside from low levels of uncleaned hot pixels, despite the poorer fit statistic which is primarily driven by the individual archival PSFs. In F606W, positive residuals near the nucleus indicate irregular structure beyond the three-component model. Conversely, the coherent arched configurations of the negative residuals is primarily due to the apparent dust lanes visible in the data image. There are similar positive residuals in the nuclear region of the F105W image as in F606W. These compact residuals could reflect circumnuclear star formation or unmodeled structure such as a nuclear bar or spiral; the current data cannot distinguish these. The remaining outer lying residuals are from several prominent foreground stars that cannot be masked during fitting because they overlap with the disk.

%We compute the stellar mass of the galaxy using the fitted magnitudes of the S\'ersic components in F105W. Assuming a fiducial mass-to-light ratio of $Y_{*} = 1.35\ M_{\odot}/L_{\odot}$ for an early-type galaxy and a Kroupa initial mass function \citep[e.g.,][]{Ricciardelli2012}, we obtain log$\frac{M_{*}}{M_{\odot}} = 9.54\pm0.30$. 

If the nuclear region does indeed contain a SMBH binary, a singular compact component in each filter encapsulates the system as a whole and not the individual SMBHs due to instrumental resolution limitations. The bulge–disk decomposition could in principle place the system on the $M_\mathrm{BH}$--$M_\mathrm{bulge}$ relation \citep{KormendyHo2013}. Given the large uncertainties in the bulge parameters (Table \ref{tab:galfit_bestfit}), we instead use the single-epoch virial mass (\S6.3).

Since the IR and UVIS images were taken $\sim$6 months apart, we caution that interpretation of nuclear colors may not accurately reflect the state of the system. The scale of AGN variability demonstrated between the epochs of our STIS spectra and the DBSP spectra of J22 ($\sim$8 months) suggests that the imaging in the nuclear region may similarly be highly variable.

%-----------------------------------------------------------------
\section{Discussion}\label{sec:discussion}

\subsection{Combined picture}

Our \hst\ observations test two largely independent predictions of
the late-inspiral binary scenario for \target. The continuum shows no coherent near-UV deficit, although Fe II template systematics preclude a quantitative limit. The kinematic test disfavors a disk-emitter at the HST epoch and shows no coherent centroid offset of the kind expected from a BLR bound to one SMBH. The broad-line properties are consistent with a normal single-AGN BLR.

Neither outcome by itself decides the binary question. The continuum non-detection is informative only against the subset of CBD parameter space that predicts a deep notch within our observed band, and recent simulations \citep{Bowen2018,dAscoli2018,Combi2022,Tiwari2025} make clear that this subset is not all of CBD parameter space. The kinematic test is informative against the disk-emitter alternative but is degenerate with a single recoiling SMBH and with transient outflows acting on a single BLR.

\subsection{Combination with post-2022 multi-wavelength evidence}

The most stringent constraints on the imminent-merger interpretation come not from any single dataset but from the cumulative evidence since the original chirp ended in early 2022.  The chirping periodicity has not continued \citep{Dotti2023,Masterson2023,Marin2023}. The \halpha\ broad-line variability tracks the continuum on light-travel timescales, not on orbital timescales \citep{Hoshi2024}. Multi-epoch VLBI shows a single unresolved compact core with no morphological evidence for a binary \citep{An2022,An2026}. No PTA detection of a burst-with-memory has been reported, although the predicted GW signal-to-noise from a J22 merger would have been marginal even for IPTA \citep{Chen2022}. The HESS collaboration reports only upper limits at TeV energies \citep{HESS2023}. Our \hst\ observations are therefore best interpreted as additional independent constraints on a candidate that, on the cumulative weight of multi-wavelength evidence, has not behaved as the original imminent-merger model predicted.

\subsection{Alternatives and caveats}

A range of alternatives to the late-inspiral binary scenario have been advanced in the post-2022 literature. \citet{Dotti2023} propose Lense--Thirring precession of an accretion disk around a single SMBH to account for the decreasing flare amplitude.  \citet{Masterson2023} and \citet{Hoshi2024} interpret the unusual hard X-ray flares and the continuum-tracking line variability as products of a variable corona or changing-look transition rather than a binary. Statistical treatment of red-noise quasar variability emphasizes that $\sim 4$--$6$-cycle ``periodicities'' suffer false-positive rates that are non-negligible \citep{Vaughan2016,Zhu2020}. \citet{He26} further find a stable hard X-ray lag, suggesting that the disk–corona geometry did not change as the X-ray flare faded.

We note that the disk-emitter alternative we test in Section~\ref{sec:lines} addresses only the broad-line shape, not the chirping light curve. None of the published alternatives requires a binary at sub-parsec separation; in this sense the imminent-merger interpretation, while not strictly excluded, currently rests on the original interpretation of the 2018--2022 light curve, which itself has plausible non-binary explanations. \citet{He26} suggest that the periodic nature of the light curve can be explained by a radiation pressure instability within the disk at the transition between a standard thin disk \citep[i.e., ][]{Shakura73} and an advection dominated accretion flow \citep[e.g., ][]{Narayan94}, which would be most pronounced at X-ray regimes leading to the prominent flaring periods. \citet{He26} posit that the decaying trend in the light curves, as well as the damping amplitudes in the optical, can be modeled by the narrow unstable zone shrinking over the course of several hundred days, although the mechanisms responsible for instigating the onset of the instability and the subsequent shrinking unstable zone are not clear. 

A tidal disruption analog, in which a star is captured in an eccentric hardening orbit and disrupted in successive pericenter passages \citep[in the spirit of][]{Cufari2022}, is qualitatively an attractive explanation for the onset of the flares given the very large $|\dot P|$ but has not been explicitly modeled for \target. Our UV-optical spectrum does not exhibit much evidence that would corroborate the hypothesis of a TDE. We detect a broad  He {\sc ii} $\lambda$1640 line (FWHM$\sim4800 \kms$, fit jointly with O {\sc iii}] $\lambda$1663), but He {\sc ii} $\lambda$4686 requires only a narrow component, and neither line is double-peaked. We also do not detect the nearby N {\sc iii} $\lambda$4640 Bowen fluorescence line typically associated with a TDE, and the N {\sc iii} 4100\AA\ Bowen feature, if present, is blended with H$\delta$ at our resolution \citep[e.g.,][]{Onori19}. Our spectrum covers [Fe {\sc vii}] $\lambda$3759, which we do not detect (3$\sigma$ EW $<$ 6.3 \AA). The stronger [Fe {\sc vii}] $\lambda\lambda$5721, 6087 and [Fe {\sc X}] $\lambda$6375 coronal lines lie redward of our coverage, and the optical spectrum of \citet{He26} shows no evidence of them, which likely would not have faded within the time between the epoch of the flares and the spectroscopic observations \citep{Yang13}. While our \hbeta\ broad line consists of multiple components, it is not the asymmetric double-peaked profile expected \citep[e.g., AT 2020zso;][]{Wevers22}. The strong observed ratio between C {\sc iii}] $\lambda$1908 and N {\sc iii}] $\lambda$1750 would be unusual for a TDE, where nitrogen is expected to be more abundant in the line-forming region of the wind due to CNO processing in the star being disrupted \citep{Parkinson20}. Lastly, the modest velocity shifts of the multiple blue-shifted C {\sc iv} components do not suggest powerful outflows associated with a recent TDE \citep[e.g., ][and references therein]{Parkinson20}. We therefore do not find compelling spectroscopic evidence of a TDE origin for the flares observed in \target.

Assuming that the system is actually a single SMBH, we compute the total black hole mass $M_{\bullet}$ using the single-epoch virial mass estimate of \citet{Shen12}: 
\begin{equation}
{\rm log}\frac{M_{\rm \bullet,\ vir}}{M_{\odot}} = a + b\ {\rm log}\frac{L}{10^{44}\  \rm erg\ s^{-1} } + c\ {\rm log}\frac{\rm FWHM}{\rm km\ s^{-1}},
\end{equation}
where $L$ is monochromatic luminosity of either an emission line or continuum, FWHM is line width, and $a, b,$ and $c$ are scaling factors dependent on choice of line/continuum region used. For the optical range of our spectrum, we use the luminosity and FWHM of the total broad \hbeta\ profile (both broad components; log$L=41.28$, FWHM$\approx5000\kms$), with coefficients $a = 1.963$, $b = 0.401$, and $c = 1.959$, and obtain log$(M_{\bullet}/M_{\odot}) = 8.1\pm0.3$. Using the core component alone would give log$(M_{\bullet}/M_{\odot}) = 7.7\pm0.3$. Our power law fit allows us to similarly compute $M_{\bullet}$ using the monochromatic luminosity at 5100 \AA; following the relation of \citet{Assef11}, $a = 0.895$, $b = 0.520$, and $c = 2.00$, obtaining log$(M_{\bullet}/M_{\odot}) = 7.05^{+0.41}_{-0.37}$. The \hbeta-based mass lies within the range log$(M_{\bullet}/M_{\odot}) = 7.58 - 8.29$ found using broad-band SED fitting by \citet{He26}. The uncertainties in both estimates are dominated by average systematic scatter in the relations added in quadrature to the 1-$\sigma$ measurement uncertainties from the spectral fitting. If the system is indeed a SMBH binary, then the virial mass estimate is derived from the BLR of the CBD. However, virial estimators typically assume a steady-state BLR; a post-flare system (either due to a binary or a TDE) may not necessarily have reached a steady state by the epoch of our spectroscopic observations, and we thus caution against a strict interpretation of the derived total mass estimate. 

The two estimates differ by $\sim$1 dex. We adopt the \hbeta-based mass, because the L5100 estimate depends on the poorly constrained host/power-law decomposition (Fig.~\ref{fig:qsofit_corner}), to compute the Eddington ratio, $\lambda_{\rm Edd} \equiv \frac{L_{\rm bol}}{L_{\rm Edd}}$. We calculate the bolometric luminosity, $L_{\rm bol}$, from $L_{\rm [OIII]}$ with a bolometric correction of 3500 \citep{heckman04}. We calculate the Eddington luminosity in the standard convention as:
\begin{equation}
L_{\rm Edd} = \frac{4\pi GM_{\bullet}m_{\rm p}c}{\sigma_{\rm T}} \approx 1.26\times10^{38}\frac{M_{\bullet}}{M_{\odot}}{\rm erg\ s^{-1}}, 
\end{equation}
where $G$ is the gravitational constant, $c$ is the speed of light, $m_{\rm p}$ is the proton mass, $\sigma_{\rm T}$ is the Thomson scattering cross-section. We obtain a modest Eddington ratio of log$\lambda_{\rm Edd} = -1.65\pm0.3$, corresponding to a mass accretion rate of $\dot{M}\sim 0.07\ M_{\odot}$/yr for a radiative efficiency of 0.1. Because $L_{[\rm O III]}$ traces the narrow-line region, averaged over $\gtrsim 10^{3}$ yr, this is a long-term average typical of Seyfert 1 nuclei. It does not constrain the accretion state at the HST epoch.

\subsection{Outlook}

If the binary interpretation of J22 were correct, coalescence should already have occurred, so post-merger signatures (e.g., a recoiling SMBH or a refilling inner disk) are now the relevant observables. UV spectroscopy at higher resolution, or with improved Fe {\sc ii} modeling, could turn our qualitative continuum result into a quantitative notch limit. Deep ground-based imaging would be more sensitive to faint tidal features than our single-orbit HST data. PTA observations remain the cleanest channel for confirming any SMBH merger.

%-----------------------------------------------------------------
\section{Summary}\label{sec:summary}

We have presented \hst/STIS UV spectroscopy and \hst/WFC3 three-band imaging of \target, the candidate coalescing SMBH binary identified by \citet{Jiang2022arxiv}. Our principal findings are:

\begin{enumerate}
    \item The combined UV+optical continuum is well described by a
    single-AGN spectral model. We find no coherent CBD-induced deficit, although Fe {\sc ii} template systematics preclude a quantitative limit. This does not refute the binary scenario, since GRMHD simulations including minidisks predict spectra similar to single-AGN templates.
    \item The comparison of \mgii\ with the broad Balmer profile disfavors a disk-emitter at the HST epoch. Both lines are single-peaked with matching widths, and the broad Mg {\sc ii} core is blueshifted relative to \hbeta\ by $\sim640 \kms$ ($\sim3\sigma$ including cross-grating systematics), with no coherent offset of the kind expected from a bound BLR. Broad \civ\ is single-peaked (with multiple components) and blueshifted, consistent with a normal accretion-disk wind, and is not used as a centroid anchor in our binary--versus--disk-emitter comparison.
    \item The host is a bulge-plus-disk (S0-like) galaxy with nuclear dust lanes; no tidal features are detected to $\mu$(F105W) $\approx$ 25.06 mag arcsec$^{-2}$. 
    \item Combined with the cumulative post-2022 multi-wavelength record, our \hst\ constraints are consistent with the emerging consensus that the imminent-merger interpretation of \target\ has not been borne out by continued monitoring, although a late-inspiral binary is not strictly excluded.
\end{enumerate}

\target\ remains a useful laboratory for the kinematic and continuum diagnostics of bound SMBH binaries. Our results illustrate the use of \mgii\ versus the broad Balmer profile, rather than \civ\ centroid shifts, as a UV-spectroscopic test of the disk-emitter alternative for candidate binaries.

\begin{acknowledgments}
The authors would like to thank Dr. Colin Burke for helpful discussions and active user-response development of {\texttt JAXSEDFit}.
We also thank Dr. Ning Jiang for helpful comments and discussions.
This material is based upon work supported by the National Science Foundation Graduate Research Fellowship Program under Grant No. DGE 21-46756. Any opinions, findings, and conclusions or recommendations expressed in this material are those of the author(s) and do not necessarily reflect the views of the National Science Foundation.
This work is based on observations made with the NASA/ESA Hubble Space Telescope, obtained at the Space Telescope Science Institute, which is operated by the Association of Universities for Research in Astronomy, Inc., under NASA contract NAS 5-26555. These observations are associated with program HST-GO-16926. Support for program HST-GO-16926 was provided by NASA through a grant from the Space Telescope Science Institute. X.L. and L.N. also acknowledge partial support from NSF grants AST-2206499.

This research has made use of the NASA/IPAC Extragalactic Database (NED), the Mikulski Archive for Space Telescopes (MAST), and observations from the SDSS, ZTF, ATLAS, CRTS, PanSTARRS, NEOWISE, \swift, and \xmm\ surveys/missions.
\end{acknowledgments}

\facilities{HST(STIS), HST(WFC3), Hale (DBSP), Swift(XRT, UVOT), XMM, Sloan, ZTF, PS1, ATLAS, CRTS, NEOWISE}

\software{calstis \citep{Hodge1998}, calwf3 \citep{Ryan16}, drizzlepac \citep{Fruchter10, Hoffmann21}, GALFIT \citep{Peng2010}, astropy \citep{astropy:2013, astropy:2018, astropy:2022}, matplotlib \citep{Hunter:2007}, numpy \citep{Harris20}, scipy \citep{2020SciPy-NMeth}, JAXQSOFIT \citep{jaxqsofit}, JAXSEDFIT \citep{jaxsedfit}, numpyro \citep{Phan19}, dsps \citep{Hearin23}, dustmaps \citep{Green18}, HASP \citep{Debes24}, AFFOGATO \citep{Nolan25}, and GALFITM \citep{Haussler22}
}

DATA AVAILABILITY: The HST/STIS and WFC3 data are available from MAST at doi:10.17909/[DOI] (program GO-16926). Reduced spectra, posterior samples from the spectral fit, and the image models are available at Zenodo [DOI].

AI tools were used to assist with refinement of the presentation of code used in the research process, mainly to aid in coding visualizations. The GAI tool used was Gemini (Google, gemini-3.7-flash and gemini-3.8-flash August -- September 2026). The scientific analysis and conclusions are the authors' own.
%-----------------------------------------------------------------
% Figure stubs - replace with actual figures when ready
%-----------------------------------------------------------------

\appendix
\section{FULL SPECTRAL-FIT RESULTS}
\begin{figure*}
\centering
 \includegraphics[width=\textwidth]{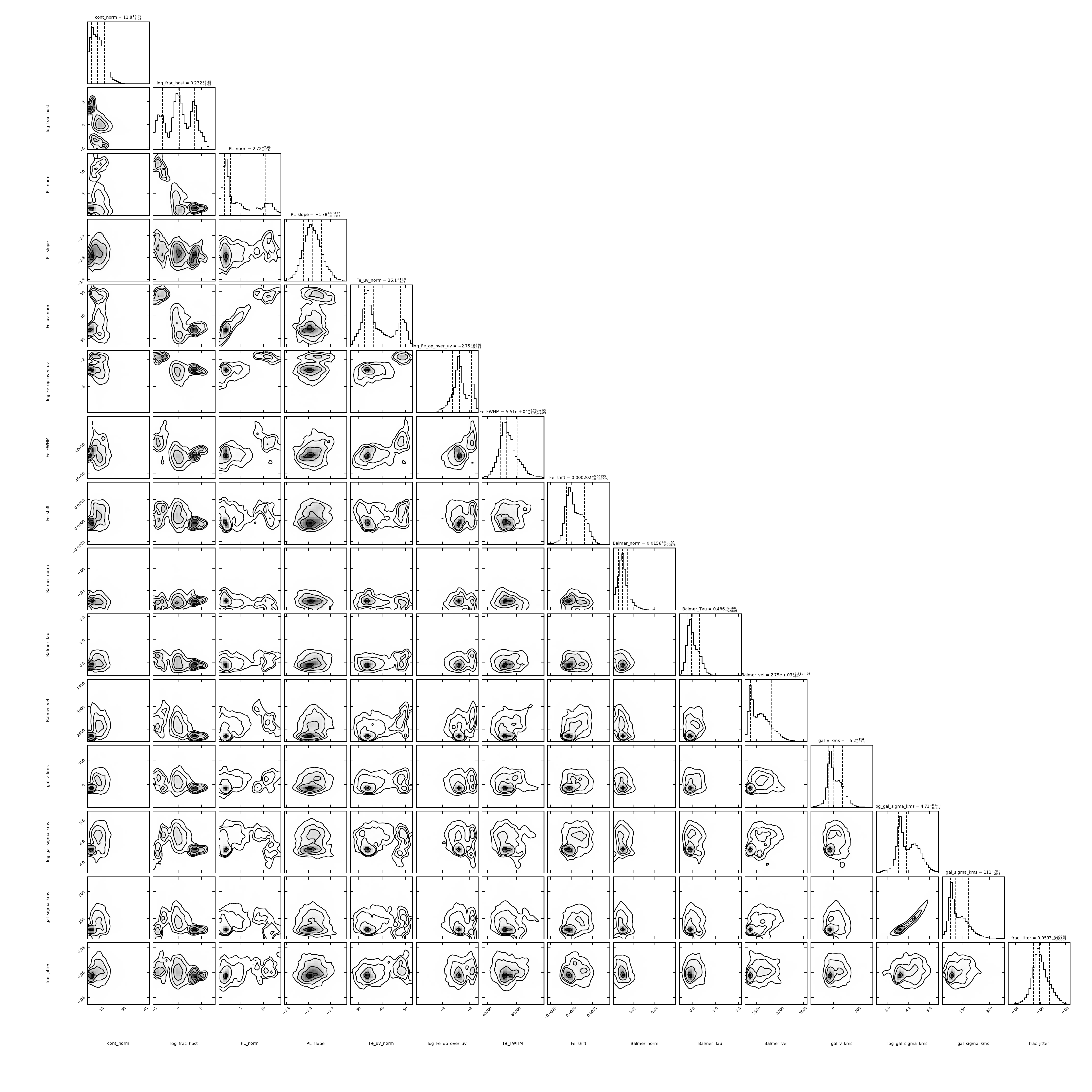}
\caption{Corner plot showing the posterior probability distributions from MCMC analysis corresponding to the best fit spectral model in Fig \ref{fig:qsofit}. The joint posterior distributions are, in general, non-Gaussian, reflecting the difficulty in constraining the complex simultaneous model fitting on a spectrum with fairly poor resolution. Notable exceptions are the power law slope and Fe{\sc ii} FWHM, whose 1D distributions appear more robustly constrained. We note that the simultaneous line decompositions appeared to be robust against changes in these component priors.  \label{fig:qsofit_corner}}
\end{figure*}

Table~\ref{tab:emi_lines_appendix} lists all emission-line components not shown in Table~\ref{tab:emi_lines}, and Figure~\ref{fig:qsofit_corner} shows the posterior distributions of the continuum parameters.

\begin{deluxetable*}{lccccr}
%\tabletypesize{\script} \tablewidth{8pt}
\tabletypesize{\normalsize}
\tablewidth{\textwidth}

\tablecaption{Additional Emission Line Properties\
\label{tab:emi_lines_appendix}}
\tablehead{ 
\colhead{Emission} & \colhead{$\lambda_{\rm vac}$}& \colhead{Luminosity}& \colhead{Velocity Offset} & \colhead{FWHM} & \colhead{EW} \\
\colhead{Line} & \colhead{\AA} & \colhead{log[erg s$^{-1}$]} & \colhead{km s$^{-1}$} & \colhead{km s$^{-1}$} & \colhead{\AA} 
}
\startdata 
H$\beta_{\rm na}$ & 4862.68 & $39.99^{+0.10}_{-0.19}$ & $274^{+32}_{-25}$ & $517^{+67}_{-52}$ & $6.9^{+1.7}_{-2.5}$ \\
\noalign{\smallskip}
[O {\sc iii}] 4959$_{\rm c}$ & 4960.3 & $40.57^{+0.02}_{-0.01}$ & $228^{+18}_{-15}$ & $706^{+0}_{-0}$ & $23.7^{+1.1}_{-0.8}$ \\
\noalign{\smallskip}
[O {\sc iii}] 5007$_{\rm c}$ & 5008.24 & $41.04^{+0.02}_{-0.01}$ & $228^{+18}_{-15}$ & $706^{+0}_{-0}$ & $72.4^{+3.3}_{-2.3}$ \\
\noalign{\smallskip}
[O {\sc iii}] 4959$_{\rm w}$ & 4960.3 & $39.81^{+0.13}_{-0.43}$ & $-1139^{+624}_{-550}$ & $2523^{+236}_{-2108}$ & $4.1^{+1.4}_{-2.6}$ \\
\noalign{\smallskip}
[O {\sc iii}] 5007$_{\rm w}$ & 5008.24 & $40.28^{+0.13}_{-0.43}$ & $-1139^{+624}_{-550}$ & $2523^{+236}_{-2108}$ & $12.6^{+4.3}_{-8.0}$ \\
\noalign{\smallskip}
H$\gamma_{\rm br}$ & 4341.68 & $40.52^{+0.11}_{-0.19}$ & $-292^{+318}_{-334}$ & $3447^{+1277}_{-623}$ & $22.7^{+6.8}_{-8.0}$ \\
\noalign{\smallskip}
H$\gamma_{\rm na}$ & 4341.68 & $39.60^{+0.23}_{-0.51}$ & $274^{+32}_{-25}$ & $517^{+67}_{-52}$ & $2.8^{+1.9}_{-1.9}$ \\
\noalign{\smallskip}
H$\delta_{\rm br}$ & 4102.89 & $40.18^{+0.14}_{-0.14}$ & $-1070^{+375}_{-406}$ & $29812^{+1162}_{-157}$ & $10.2^{+3.8}_{-2.8}$ \\
\noalign{\smallskip}
H$\delta_{\rm na}$ & 4102.89 & $39.25^{+0.34}_{-1.04}$ & $274^{+32}_{-25}$ & $517^{+67}_{-52}$ & $1.2^{+1.4}_{-1.1}$ \\
\noalign{\smallskip}
[O {\sc ii}] 3728 & 3728.48 & $40.26^{+0.17}_{-0.32}$ & $274^{+32}_{-25}$ & $517^{+67}_{-52}$ & $12.8^{+6.1}_{-6.7}$ \\
\noalign{\smallskip}
[Ne {\sc v}] 3426 & 3426.84 & $39.97^{+0.22}_{-0.86}$ & $143^{+200}_{-131}$ & $758^{+173}_{-153}$ & $3.5^{+2.4}_{-3.0}$ \\
\noalign{\smallskip}
Mg {\sc ii}$_{\rm na}$ & 2798.75 & $39.16^{+0.45}_{-0.44}$ & $274^{+32}_{-25}$ & $517^{+67}_{-52}$ & $0.4^{+0.7}_{-0.2}$ \\
\noalign{\smallskip}
C {\sc iii}]$_{\rm na}$ & 1908.73 & $40.34^{+0.10}_{-0.13}$ & $228^{+18}_{-15}$ & $706^{+0}_{-0}$ & $7.9^{+2.1}_{-2.1}$ \\
\noalign{\smallskip}
Si {\sc iii}] 1892 & 1892.03 & $39.14^{+0.50}_{-0.53}$ & $228^{+18}_{-15}$ & $706^{+0}_{-0}$ & $0.5^{+1.2}_{-0.4}$ \\
\noalign{\smallskip}
Al {\sc iii}1857 & 1857.4 & $39.01^{+0.37}_{-0.41}$ & $228^{+18}_{-15}$ & $706^{+0}_{-0}$ & $0.4^{+0.5}_{-0.2}$ \\
\noalign{\smallskip}
Si {\sc ii}1816 & 1816.98 & $39.60^{+0.19}_{-0.40}$ & $228^{+18}_{-15}$ & $706^{+0}_{-0}$ & $1.5^{+0.8}_{-0.9}$ \\
\noalign{\smallskip}
N {\sc iii}] 1750 & 1750.26 & $39.79^{+0.26}_{-0.29}$ & $228^{+18}_{-15}$ & $706^{+0}_{-0}$ & $2.3^{+1.9}_{-1.1}$ \\
\noalign{\smallskip}
N {\sc iv} 1718 & 1718.55 & $39.73^{+0.16}_{-0.23}$ & $228^{+18}_{-15}$ & $706^{+0}_{-0}$ & $2.0^{+0.9}_{-0.8}$ \\
\noalign{\smallskip}
C {\sc iv}$_{\rm na}$ & 1549.06 & $40.16^{+0.21}_{-0.58}$ & $228^{+18}_{-15}$ & $706^{+0}_{-0}$ & $5.1^{+3.1}_{-3.9}$ \\
\noalign{\smallskip}
O {\sc iii}] 1663 & 1663.48 & $39.22^{+0.35}_{-0.45}$ & $228^{+18}_{-15}$ & $706^{+0}_{-0}$ & $0.6^{+0.7}_{-0.4}$ \\
\noalign{\smallskip}
O {\sc iii}] 1663$_{\rm br}$ & 1663.48 & $40.67^{+0.09}_{-0.39}$ & $-119^{+1452}_{-1461}$ & $4827^{+1710}_{-1620}$ & $16.7^{+4.1}_{-10.0}$ \\
\noalign{\smallskip}
He {\sc ii} 1640 & 1640.42 & $39.40^{+0.38}_{-0.65}$ & $228^{+18}_{-15}$ & $706^{+0}_{-0}$ & $0.9^{+1.2}_{-0.7}$ \\
\noalign{\smallskip}
He{ \sc ii} 1640$_{\rm br}$ & 1640.42 & $40.76^{+0.17}_{-0.32}$ & $-119^{+1452}_{-1461}$ & $4827^{+1710}_{-1620}$ & $19.9^{+9.6}_{-10.4}$ \\
\noalign{\smallskip}
[O {\sc ii}] 3726 & 3726.03 & $39.83^{+0.25}_{-0.46}$ & $274^{+32}_{-25}$ & $517^{+67}_{-52}$ & $4.7^{+3.7}_{-3.1}$ \\
\noalign{\smallskip}
[O{ \sc ii}] 3729 & 3728.82 & $39.83^{+0.25}_{-0.46}$ & $274^{+32}_{-25}$ & $517^{+67}_{-52}$ & $4.7^{+3.7}_{-3.1}$ \\
\noalign{\smallskip}
[Ne {\sc iii}] 3869 & 3869.86 & $40.27^{+0.09}_{-0.06}$ & $228^{+18}_{-15}$ & $706^{+0}_{-0}$ & $11.6^{+2.7}_{-1.5}$ \\
\noalign{\smallskip}
[Ne {\sc iii}] 3968 & 3968.59 & $40.01^{+0.12}_{-0.15}$ & $228^{+18}_{-15}$ & $706^{+0}_{-0}$ & $7.2^{+2.4}_{-2.1}$ \\
\noalign{\smallskip}
[O {\sc iii}] 4363 & 4364.44 & $39.75^{+0.19}_{-0.22}$ & $228^{+18}_{-15}$ & $706^{+0}_{-0}$ & $3.5^{+2.0}_{-1.4}$ \\
\noalign{\smallskip}
He {\sc ii} 4686 & 4687.02 & $39.69^{+0.20}_{-0.33}$ & $228^{+18}_{-15}$ & $706^{+0}_{-0}$ & $3.1^{+1.8}_{-1.6}$ \\
\noalign{\smallskip}
[Ne {\sc v}] 3346 & 3346.79 & $39.88^{+0.18}_{-0.36}$ & $143^{+200}_{-131}$ & $758^{+173}_{-153}$ & $2.6^{+1.4}_{-1.5}$
\enddata
\tablecomments{ 
Extension of Table~\ref{tab:emi_lines}. Luminosity is reported in log units of erg s$^{-1}$, while the velocity offset from the vacuum central wavelength and the full width at half max (FWHM) are in units of km s$^{-1}$. Broad and narrow line components are denoted as br and na, respectively. Values are quoted as the parameter's median value across draws. Errors are quoted at the 1-$\sigma$ level from Markov Chain Monte Carlo simulations. Note that the velocity offsets shown here have not been adjusted based on the systemic velocity measured for the core \oiii\ line ($+228^{+18}_{-15}$$\kms$). Narrow lines are tied in two groups (Balmer, Mg {\sc ii} and [O {\sc ii}]; [O {\sc iii}] and the UV narrow lines), each sharing one velocity and one width. The [O {\sc iii}]-group width converged to the prior bound.
}
\end{deluxetable*}

\bibliographystyle{aasjournal}
\bibliography{j1430}
%\bibliography{allrefs}

\end{document}